\documentclass[showpacs,amsmath,amssymb,aps,twocolumn,pra]{revtex4-1}

\usepackage{graphicx}% Include figure files
\usepackage{dcolumn}% Align table columns on decimal point
\usepackage{bm}% bold math

\usepackage[usenames]{color}

\usepackage{pgfplots}
\usepgfplotslibrary{external}
\tikzsetexternalprefix{Figures/Tikz/}

\usetikzlibrary{arrows,shapes,backgrounds,calc,
    positioning,
    intersections}

\newcommand{\ket}[1]{\left| #1 \right>} % for Dirac bras
\newcommand{\LLLket}[1]{\left\| #1 \right>} % for LLL kets

\newcommand{\psif}{\psi_{\mathrm{f}}}

\newcommand{\rr}{\mathbf{r}}

\newcommand{\dd}{\mathrm{d}}

\begin{document}

\title{Creating fractional quantum Hall states with atomic clusters\\ using light-assisted insertion of angular momentum}

\author{Junyi Zhang\footnote{Current address: Department of Physics, Princeton University, Princeton, 08544, New Jersey, USA}}

 \altaffiliation[Current address: ]{Department of Physics, Princeton University, Princeton, 08544, New Jersey, USA.}

\author{J\'er\^ome Beugnon}

\author{Sylvain Nascimbene}

\email{sylvain.nascimbene@lkb.ens.fr}

\affiliation{%
Laboratoire Kastler Brossel, Coll\`ege de
France, ENS-PSL Research University, CNRS, UPMC-Sorbonne Universit\'es, 11 place Marcelin Berthelot, 75005 Paris, France
}%

\date{\today}

\begin{abstract} 
We describe a protocol to prepare clusters of ultracold bosonic atoms in strongly-interacting states reminiscent of fractional quantum Hall states. Our scheme consists in injecting a controlled amount of angular momentum to an atomic gas using Raman transitions carrying orbital angular momentum. By injecting one unit of angular momentum per atom, one realizes a single-vortex state, which is well described by mean field theory for large enough particle numbers. We also present schemes to realize fractional quantum Hall states, namely the bosonic Laughlin and Moore-Read states. We investigate the requirements for adiabatic nucleation of a such topological states, in particular comparing linear Landau-Zener ramps and arbitrary ramps obtained from optimized control methods. We also show that this protocol requires excellent control over the isotropic character of the trapping potential. 
\end{abstract}

\maketitle

\section{Introduction}
Ultracold atom experiments provide unique playgrounds for investigating complex states of matter in a controlled environment, such as strongly-interacting Fermi gases, low-dimensional states of matter or lattice quantum systems \cite{bloch2008many}. The effect of a magnetic field on charged quantum many-body systems leads to a wealth of interesting states of matter, such as integer and fractional quantum Hall states. Exploring this field with ultracold atoms requires creating an artificial magnetic field that mimics the Lorentz force acting on charged particles. In the recent years the simulation of such gauge fields was extensively developed along several directions, including setting gases in rotation, dressing atoms with laser fields, and using time-modulated optical lattices \cite{cooper2008rapidly,dalibard2011colloquium,galitski2013spin,goldman2014light}. 

The physical behavior of atomic gases in the presence of an artificial magnetic field was studied with Bose-Einstein condensates, for high filling factors $\nu=N/N_v$ corresponding to a number of flux quanta $N_v$ much less than the particle number $N$. In that regime, the gauge field leads to the appearance of $N_v$ quantized vortices piercing the Bose-Einstein condensate \cite{madison2000vortex,abo2001observation,zwierlein2005vortices,lin2009synthetic}. The quantum Hall regime was reached using rotating gases with large filling factors \cite{schweikhard2004rapidly,bretin2004fast}, but the strongly correlated regime is expected for fillings $\nu\sim1$, which seems realistic to reach in experiments with small atomic samples only  \cite{cooper2008rapidly}.

Here, we describe an experimental scheme for preparing a system of a few atoms in strongly-correlated states \cite{popp2004adiabatic,roncaglia2010pfaffian,roncaglia2011rotating,julia2011strongly,ramachandhran2013emergence,grass2013fractional},  similar to the ones associated with the fractional quantum Hall effect (FQHE) \cite{cooper1999composite,wilkin2000condensation,cooper2001quantum,paredes20011,sinova2002quantum,mueller2002two,regnault2003quantum,popp2004adiabatic,barberan2006ordered,baur2008stirring,julia2011strongly}. We propose to transfer a controlled amount of angular momentum using Raman transitions, making use of Laguerre-Gauss Raman beams to transfer orbital angular momentum to the atoms. Our scheme allows transferring a given (integer) number $p$ of angular momentum quanta per atom, leading to a non-trivial dynamics in the lowest Landau level (LLL). In particular we propose a method to adiabatically prepare the LLL ground state of fixed angular momentum $L=pN$ (we set $\hbar=1$). We discuss the examples of the one-vortex $L=N$ state \cite{wilkin1998attractive,bertsch1999yrast,smith2000exact,jackson2000analytical}, as well as paradigmatic FQHE states, the Laughlin and Moore-Read states \cite{laughlin1983anomalous,moore1991nonabelions}, occurring for angular momenta $L=N(N-1)$ and $L=N(N-2)/2$, respectively. We also discuss the requirements for adiabaticity, as well as shortcuts to adiabaticity using optimized variations of the system parameters. Our study is restricted to bosonic atoms but it could be transposed to fermions straightforwardly.

\section{Description of the scheme}

%%%%%%%%%%%%%%%%%%%%%%%%%%%%%%%%%%%%%%%%%%%%%
% Figure Scheme
%%%%%%%%%%%%%%%%%%%%%%%%%%%%%%%%%%%%%%%%%%%%%

\begin{figure}
\includegraphics[width=\linewidth]{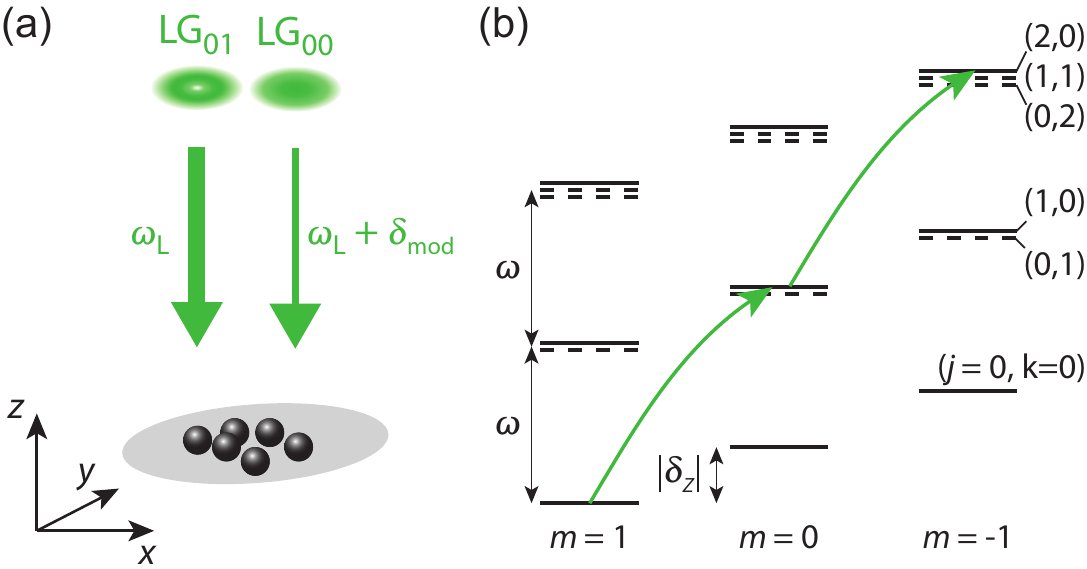}
%\vspace{-6mm}
\caption{\label{Fig_scheme}
(a) Scheme of the experiment, based on a cluster of bosonic atoms, strongly confined along $z$, and subjected along $x$ and $y$ to an isotropic harmonic trap of frequency $\omega$. Laguerre-Gauss laser beams are sent on the atomic sample to drive Raman transitions transferring orbital angular momentum.
(b) Scheme of the involved energy levels $\LLLket{m;j,k}$, indexed by the spin projection $m$ and the quanta $j,k$ of motion along $x$ and $y$, in the presence of an isotropic harmonic trap (`left-right' basis). Angular momentum is injected using Raman transitions, which couple $\LLLket{m;j,k}$ to $\LLLket{m-1;j+1,k}$ states. The trap isotropy ensures that orbital angular momentum is conserved, and $k\neq 0$ states remain unpopulated. The Raman transitions are resonant for a detuning $\delta_{\mathrm{mod}}$ equal to the difference between  the trap frequency $\omega$ and the Zeeman detuning $\delta_Z$ (negative on the figure). On the example pictured here of a spin $F=1$, an adiabatic ramp of the detuning $\delta_{\mathrm{mod}}$ across resonance leads to a transfer of two units of angular momentum per atom.
}
\end{figure}
%%%%%%%%%%%%%%%%%%%%%%%%%%%%%%%%%%%%%%%%%%%%%

%
We consider a cluster of bosonic atoms of (pseudo-)spin $F$, strongly confined along the spatial direction $z$, leading to quasi-2D dynamics in the $x-y$ plane. An additional harmonic confinement is produced along $x$ and $y$, of angular frequency $\omega$ -- the trap being assumed perfectly isotropic. The harmonic motion of single-particle eigenstates can be quantized in the `left-right' basis as $\ket{j,k}$, $j,k\in\mathbb{N}$, with an energy $\hbar\omega(j+k+1)$ and orbital angular momentum projection $l$ along $z$, where $l=j-k$. The family of states $\ket{j=l,k=0}$ of maximal angular momentum forms a basis of the lowest Landau level (LLL) of charged particles in a magnetic field, of cyclotron frequency $2\omega$. Including the internal spin degree of freedom, we write single-particle eigenstates in the basis $\LLLket{m;j,k}$, where $-F\leq m\leq F$ denotes  the spin projection along $z$.

%We are interested in this paper in the preparation of strongly-correlated states of atomic ensembles within the LLL, bearing strong similarities with fractional quantum Hall states observed in two-dimensional electron gases. 

%The atoms are confined in the intensity node of a blue-detuned optical dipole trap prepared in a Laguerre-Gauss LG$_{01}$ mode. The extent of the atomic ensemble is assumed to be much smaller that the dipole trap beam waist, such that the $x-y$ confinement can be considered as harmonic, of trapping frequency $\omega$ -- the trap being assumed perfectly isotropic. 

The proposed scheme is sketched in Fig.\,\ref{Fig_scheme}. We assume an initial state composed of $N$ atoms, spin-polarized and in the motional ground state $\LLLket{m=F;j=0,k=0}$. A bias magnetic field provides a spin quantization axis and lifts the spin degeneracy, modeled by a Zeeman energy $E_Z=\delta_Z m$. 
The transfer of angular momentum is provided by two-photon Raman transitions involving two Laguerre-Gauss laser beams of modes LG$_{01}$ and LG$_{00}$, of frequency difference  $\delta_{\mathrm{mod}}$. In the regime of large bias magnetic fields, one may use a rotating wave approximation, leading to a simple form for the single-particle Hamiltonian:
\begin{align*}
\hat H_1=&(\delta_Z+\delta_{\mathrm{mod}})\hat F_z+\omega(\hat a_j^\dagger\hat a_j+\hat a_k^\dagger\hat a_k+1)\\
&+\frac{\Omega}{2}\left[\hat F_-(\hat a_j^\dagger+\hat a_k)+\hat F_+(\hat a_j+\hat a_k^\dagger)\right],
\end{align*}
where $\hat a_j$ ($\hat a_k$) annihilate one right- (left-)handed quantum labeled by the integer $j$ ($k$, respectively) and $\Omega$ denotes the Rabi frequency of the Raman coupling.

We aim at driving the system in the LLL, i.e. with only $k=0$ states populated. The couplings  $\hat F_-\hat a_j^\dagger$ and $\hat F_+\hat a_k^\dagger$ are resonant for modulation frequencies $\delta_{\mathrm{mod}}=-\delta_Z+\omega$ and $\delta_{\mathrm{mod}}=-\delta_Z-\omega$, respectively. The dynamics in the LLL is induced by working around the resonance of processes $\LLLket{m;j,k}\rightarrow\LLLket{m-1;j+1,k}$, induced by the coupling $\hat F_-\hat a_j^\dagger$. We thus introduce the detuning $\delta\equiv\delta_{\mathrm{mod}}+\delta_Z-\omega$ as the control parameter for the angular momentum injection. Under the assumptions $|\delta|,\Omega\ll\omega$, the other transitions  $\LLLket{m;j,k}\rightarrow\LLLket{m+1;j,k+1}$ remain off resonant, and the single-particle Hamiltonian can be restricted to (up to a constant)
%
%%As the system is initiated in the state  $\LLLket{m=F;j=0,k=0}$, we prevent the population of $k\neq0$ state by keeping the action of the coupling $\hat F_\hat a_k^\dagger$ off resonant.
%
%
%The transition $\LLLket{m;j,k}\rightarrow\LLLket{m-1;j+1,k}$ is resonant for a frequency difference $\delta_{\mathrm{mod}}=-\delta_Z+\omega$, respectively. By driving the system close to this resonance, we forbid the 
%As the dynamics starts in the state $\LLLket{m=F;j=0,k=0}$, it is clear that the laser coupling will not drive atoms in the states of left-handed quanta $k\geq 1$, i.e. the dynamics is restricted to the LLL $k=0$. The transition $\LLLket{m;j,k}\rightarrow\LLLket{m-1;j+1,k}$ is resonant for a frequency difference $\delta_{\mathrm{mod}}=-\delta_Z+\omega$. We thus introduce  the resonance detuning $\delta\equiv\delta_{\mathrm{mod}}+\delta_Z-\omega$ as the control parameter for the angular momentum injection, leading to a simplified single particle Hamiltonian within the LLL (up to a constant)
\[
\hat H_1'=\delta\hat F_z+\frac{\Omega}{2}\left(\hat F_-\hat a_j^\dagger+\hat F_+\hat a_j\right),
\]
for which the LLL is stable. 
 When ramping $\delta$ slowly across the resonance $\delta=0$, one expects, in the absence of interactions, to adiabatically transfer all atoms in the state $\LLLket{m=-F;j=2F,k=0}$. A residual trap anisotropy would induce an additional coupling to $k\neq0$ states, that we discuss at the end of the article.

In this process, we expect interactions to play a crucial role. Indeed, in the absence of interactions the many-body state with all atoms in $\LLLket{m=-F;j=2F,k=0}$  is degenerate with all states of $N$ particles occupying the states $\LLLket{m=-F;j=l_i,k=0}$ ($1\leq l_i\leq N$), provided the energy conservation is fulfilled, i.e. $\sum_il_i=2FN$. As a result, interactions play a non-perturbative role, and the true many-body ground state occurring in the presence of interactions is likely to be strongly-correlated, i.e. not captured by a mean-field analysis.  We assume in the following that interactions can be described as contact, spin-independent interactions of scattering length $a$, leading to a coupling constant $\tilde g=\sqrt{8\pi}a/l_z$ describing collisions in a quasi-2D geometry ($l_z$ denotes the extent of the wave-function along the strong confinement axis $z$) \cite{hadzibabic2011two}. Importantly, elastic contact interactions conserve energy and orbital angular momentum, which ensures that the LLL subspace $k=0$ is stable under collisions.

In the following, assuming the dynamics to be restricted to the LLL, we write many-body wavefunctions  as  $\psi(z_i)=P(z_i)\exp(-\sum_i |z_i|^2/2a_{\mathrm{ho}}^2)$, where $z_i$ is the complex coordinate of particle $i$ in the $x-y$ plane, and $P(z_i)$ is a polynomial function of the variables $z_i$ ($1\leq i \leq N$) \cite{cooper2008rapidly}. All lengths are expressed in units of the ground state extent $a_{\mathrm{ho}}=\sqrt{\hbar/m\omega}$, and the gaussian factor is further omitted in many-body wavefunctions. Energies are  written in units of the trap frequency $\omega$.

\section{One-vortex state}
\begin{figure*}
\includegraphics[width=\linewidth]{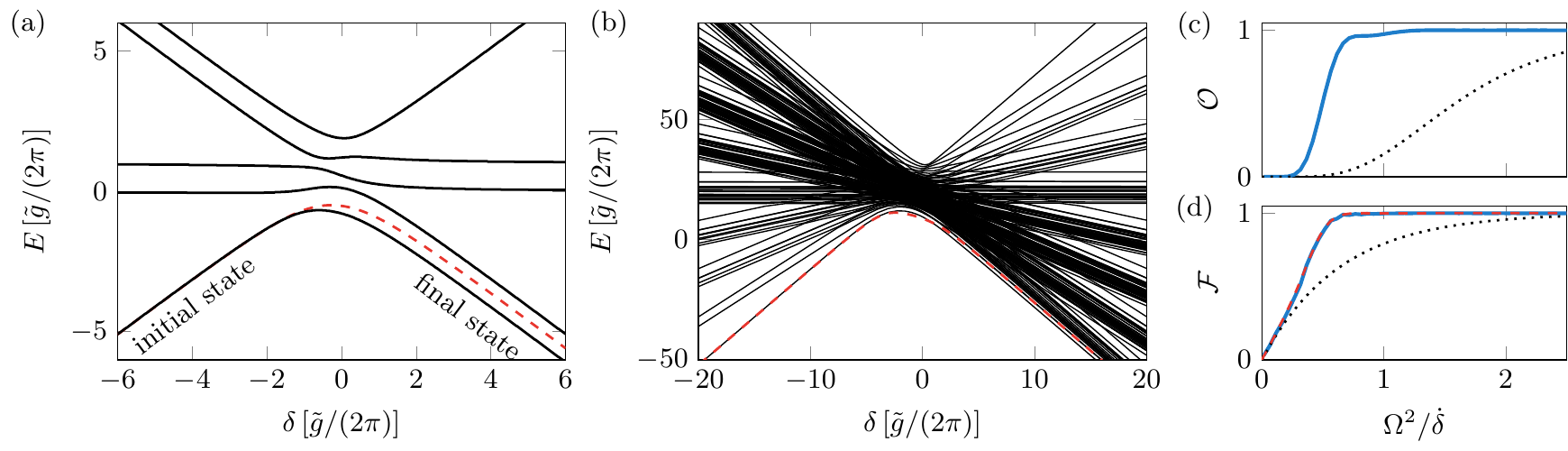}
\vspace{-7mm}
\caption{\label{Fig_L_eq_N}
(a,b) Energy levels $E$ (black lines) as a function of the detuning $\delta$, for $N=2$ (a) and $N=8$ (b) particles of spin $F=1/2$, which corresponds to a transfer of 1 unit of angular momentum per atom. The Raman coupling amplitude is set to $\Omega=\tilde g/(2\pi)$. The system is prepared at rest, spin-polarized in $m=F$, corresponding to the many-body ground state for $\delta<0$. The adiabatic detuning ramp  connects it to the one-vortex state $\ket{\psi_{\mathrm{v}}}$, without any level crossing. The red dashed lines correspond to the lowest energy of stationary states of the mean-field non-linear Schr\"odinger equation (\ref{eq_NLSE}).
(c,d) Overlap $\mathcal{O}$ (c) and fidelity $\mathcal{F}$ (d) between the one-vortex state $\ket{\psi_{\mathrm{v}}}$ and the state prepared for a ramp of finite speed $\dot\delta$, as a function of the dimensionless parameter $\Omega^2/\dot\delta$ (blue solid lines). The overlap and fidelity are defined according to  eqs.\,(\ref{eq_overlap}) and (\ref{eq_fidelity}), respectively. The calculation is performed for $N=8$ particles, and the result is compared with the Landau-Zener prediction expected for non-interacting systems (dotted lines). The calculated fidelity coincides with the one obtained from a mean-field description (dashed red line). }
\end{figure*}

We first consider the simplest case of a spin $F=1/2$, for which a single unit of angular momentum is transferred per atom, leading to the ground state of the LLL of angular momentum $L=N$. The structure of levels is illustrated in Fig.\,\ref{Fig_L_eq_N}a on the case $N=2$. The system is prepared with both atoms in the state $\LLLket{m=1/2;j=0,k=0}$, and the detuning $\delta$
 is ramped across 0  in the positive direction. An adiabatic following of the lowest energy state leads to both atoms polarized in the $m=-1/2$ state, with $L=2$. In the absence of interactions, one ends into a degenerate subspace spanned by the vectors $\ket{\psi_1}\propto\hat c^{\dagger2}_{-1/2;1,0}\ket{0}$ and  $\ket{\psi_2}=\hat c^{\dagger}_{-1/2;2,0}\hat c^{\dagger}_{-1/2;0,0}\ket{0}$, where $\hat c^{\dagger}_{m;j,k}$ creates a particle in the state $\LLLket{m;j,k}$.  Interactions lift this degeneracy, leading to a ground state $\ket{\psi_{\mathrm{v}}}=(\ket{\psi_1}-\ket{\psi_2})/\sqrt{2}$, separated in energy from the first excited state by $\tilde g/(2\pi)$.

%%%%%%%%%%%%%%%%%%%%%%%%%%%%%%%%%%%%%%%%%%%%%
% Figure L=N
%%%%%%%%%%%%%%%%%%%%%%%%%%%%%%%%%%%%%%%%%%%%%

\newlength{\lenA}
\setlength{\lenA}{9.8mm}

%%%%%%%%%%%%%%%%%%%%%%%%%%%%%%%%%%%%%%%%%%%%%

A similar structure occurs for larger particle numbers. The subspace $L=N$ of the LLL features a unique ground state, of wavefunction
\begin{equation}
\psi_{\mathrm{v}}(z_i)=\prod_{1\leq i\leq N}(z_i-z_c),\quad z_c=\frac{1}{N}\sum_{1\leq i\leq N}z_i,\label{eq_state_L_eq_N}
\end{equation}
and energy $E= N(N-2)/2$ \cite{wilkin1998attractive,bertsch1999yrast,smith2000exact,jackson2000analytical}. In the limit of large particle numbers, the fluctuations of the center-of-mass position $z_c$ decrease to 0, leading to a Bose-Einstein condensate of wavefunction $\psi_{\mathrm{v}}(z_i)=\prod_{1\leq i\leq N}z_i$, with one vortex at the trap bottom \cite{lieb2006derivation}. 
We plot in Fig.\,\ref{Fig_L_eq_N}b the energy levels of the many-body Hamiltonian as a function of the detuning $\delta$ for a system of $N=8$ particles, which correspond to a Hilbert space of dimension 185. We observe that the ground state remains gapped; thus an adiabatic ramp of the detuning $\delta$ should lead to the ground state of the LLL with $L=N$, spin-polarized in the state $\ket{F,m=-F}$. We confirm this result numerically by solving the Schr\"odinger equation with the many-body Hamiltonian, for a linear detuning ramp of speed $\dot\delta$, connecting initial and final values $\delta=-20\,\tilde g/(2\pi)$ and $\delta=20\,\tilde g/(2\pi)$, respectively. We compare the quantum state $\ket{\psif}$ obtained numerically with the ground state (\ref{eq_state_L_eq_N}), by calculating the many-body overlap 
\begin{equation}\label{eq_overlap}
\mathcal{O}\equiv\left|\left<\psif\middle|\psi_{\mathrm{v}}\right>\right|^2. 
\end{equation}
We observe that the overlap tends to 1 in the slow ramp limit. This calculation is compared with the overlap expected for non-interacting particles, given by the  Landau-Zener (LZ) formula
\begin{equation}\label{eq_LZ}
\mathcal{O}=\left[1-\exp\left(-2\pi\frac{\Omega^2}{\dd\delta/\dd t}\right)\right]^N.
\end{equation}
The larger overlap values obtained for interacting particles is reminiscent of the behavior observed with Bose-Einstein condensates into optical lattices \cite{morsch2001bloch} or coupled 1D Bose liquids \cite{chen2011many}.

In the limit of large particle numbers, we expect the system to be well described as a Bose-Einstein condensate of wavefunction $\psi=\alpha\ket{a}+\beta\ket{b}$,  occupying the two modes $\ket{a}\equiv\LLLket{1/2;0,0}$ and $\ket{b}\equiv\LLLket{-1/2;1,0}$. The time evolution of the BEC wavefunction is governed by the non-linear Schr\"odinger equation (NLSE) \cite{wu2000nonlinear,zobay2000time}
\begin{align}
&i\,\dot a=\frac{\delta}{2}a+\frac{N-1}{2}\left(|a|^2+\frac{|b|^2}{4}\right)a+\frac{\Omega}{2}b,\notag\\
&i\,\dot b=-\frac{\delta}{2}b+\frac{N-1}{2}\left(\frac{|a|^2}{4}+\frac{|b|^2}{2}\right)b+\frac{\Omega}{2}a,\notag\\
&|a|^2+|b|^2=1.\label{eq_NLSE}
\end{align}
We plot in Fig.\ref{Fig_L_eq_N}b the lowest energy $E$ associated with stationnary solutions of (\ref{eq_NLSE}) for $N=8$, which is close to the actual ground state energy for all detuning values. The many-body overlap $\mathcal{O}$ is not suited for comparing the calculations performed on the many-body wavefunctions and the mean-field description. We thus introduce the fidelity 
\begin{equation}\label{eq_fidelity}
\mathcal{F}=\mathrm{Tr}\sqrt{\sqrt{\hat\rho_{\mathrm{f}}}\hat\rho_{\mathrm{v}}\sqrt{\hat\rho_{\mathrm{f}}}},
\end{equation}
defined from the single-particle density matrices $\hat\rho_{\mathrm{v}}$ and $\hat\rho_{\mathrm{f}}$ associated with the states $\ket{\psi_{\mathrm{v}}}$ and $\ket{\psif}$, respectively \cite{uhlmann1976transition,jozsa1994fidelity}. As shown in Fig.\,\ref{Fig_L_eq_N}d, the fidelities calculated with the full many-body system and from the NLSE are in good agreement, for all values of the detuning ramp speed. 

\section{Laughlin state}
\begin{figure}
\includegraphics[width=\linewidth]{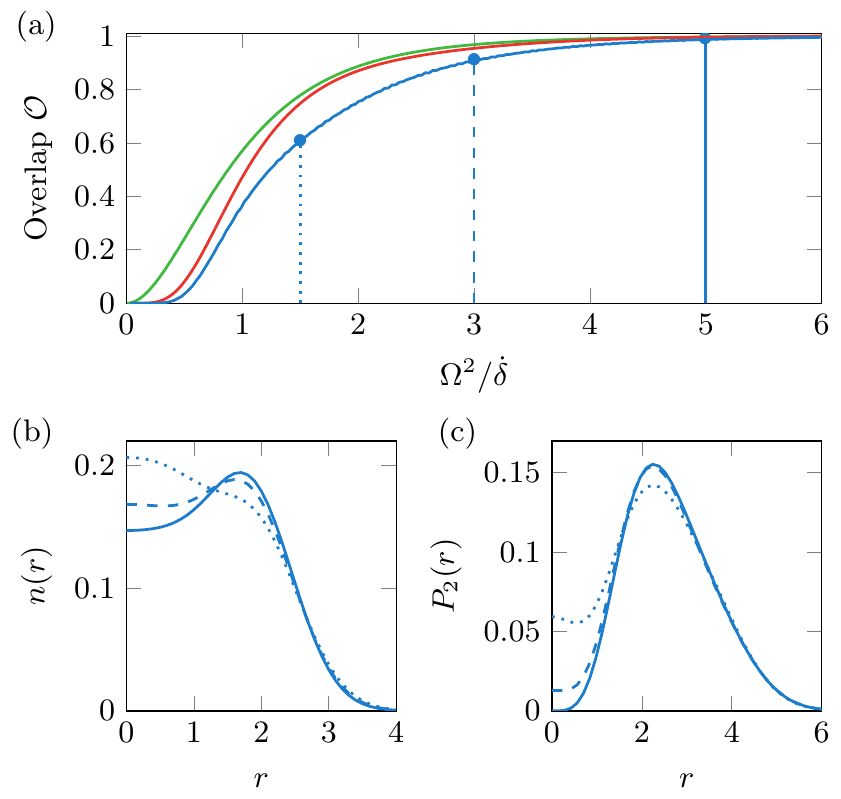}
\vspace{-6mm}
\caption{\label{Fig_Laughlin}
(a) Overlap between the Laughlin state and the state reached for a detuning ramp of speed $\dot\delta$, for particle numbers $N=2,3,4$ (green, red, blue lines, respectively), and a coupling $\Omega=\tilde g/(2\pi)$. The vertical lines indicate the 3 ramp speeds considered in (b) and (c).
(b) Atom density corresponding to the state reached for ramp speeds $\Omega^2/\dot\delta=1.5,3,5$ (dotted, dashed, solid blue lines, respectively) and $N=4$. The atom density of the $N=4$ Laughlin state is indistinguishable from the solid line.
(c) Density probability $P_2(r)$ of the inter-particle distance $r$ calculated for the three ramp speeds.
}
\end{figure}

We now discuss the realization of strongly-correlated states, which are bosonic analogs of fractional quantum Hall states observed with 2D electron gases. For contact interactions, the ground state of the LLL with $N$ particles, and total angular momentum $L=N(N-1)$, is exactly given by the Laughlin state at filling $\nu=1/2$, of wavefunction
\begin{equation}\label{eq_Laughlin}
\psi_{\mathrm{L}}=\prod_{1\leq i<j \leq N}(z_i-z_j)^2.
\end{equation}
The Laughlin state can be reached in our scheme for an adiabatic transfer of $N-1$ units of angular momentum per atom, i.e. for a pseudo-spin $F=(N-1)/2$. We performed a numerical study of the quantum state evolution during the ramp detuning, for particle numbers $N=2,3,4$, corresponding to Hilbert spaces of dimensions 5, 61, 1417, respectively. The overlap between the Laughlin state and the state reached after the detuning ramp is calculated numerically for $N=2,3,4$ particles and for different ramp speeds. As shown in Fig.\,\ref{Fig_Laughlin}a, a high overlap $\mathcal{O}>0.99$ can be obtained for ramp speeds $\dot\delta<0.2\,\Omega^2$.

%%%%%%%%%%%%%%%%%%%%%%%%%%%%%%%%%%%%%%%%%%%%%
% Figure Laughlin
%%%%%%%%%%%%%%%%%%%%%%%%%%%%%%%%%%%%%%%%%%%%%

%%%%%%%%%%%%%%%%%%%%%%%%%%%%%%%%%%%%%%%%%%%%%

As the many-body overlap cannot be accessed in experiments, we also calculated  the atom density profile and the density-density correlation function, which give a more physical insight on the prepared quantum state. The density profile  $n(r)$ should feature a plateau at its center, reminiscent of the incompressibility of the Laughlin state in the thermodynamic limit.  As shown in Fig.\,\ref{Fig_Laughlin}b on the case $N=4$, the prepared state exhibits a plateau in the middle of the trap for ramp speeds $\dot\delta\lesssim 0.3\,\Omega^2$. The Laughlin state also exhibits a strong particle anti-bunching
%, i.e. the second-order correlation function $G_2(\mathbf{r},\mathbf{r}')$ vanishes for $\mathbf{r}\rightarrow\mathbf{r}'$. 
that can be revealed from the density probability $P_2(r)$ of the inter-particle distance, defined as
\begin{align}
&2\pi r P_2(r)=\label{eq_P2}\\
&\frac{1}{N}\left<\int\dd\rr_1\dd\rr_2\,\hat\psi^\dagger(\rr_1)\hat\psi^\dagger(\rr_2)\hat\psi(\rr_2)\hat\psi(\rr_1)\delta(||\rr_2-\rr_1||-r)\right>.\nonumber
\end{align}
 As shown in Fig.\,\ref{Fig_Laughlin}c, a strong anti-bunching appears for ramp speeds $\dot\delta\lesssim 0.3\,\Omega^2$. For the particle numbers investigated here, the  physical characteristics of the Laughlin state thus appear for ramp durations comparable to the ones required for the adiabatic spin flip of a single particle.

%%%%%%%%%%%%%%%%%%%%%%%%%%%%%%%%%%%%%%%%%%%%%
% Figure Optimized ramp
%%%%%%%%%%%%%%%%%%%%%%%%%%%%%%%%%%%%%%%%%%%%%

\begin{figure*}
\includegraphics[width=\linewidth]{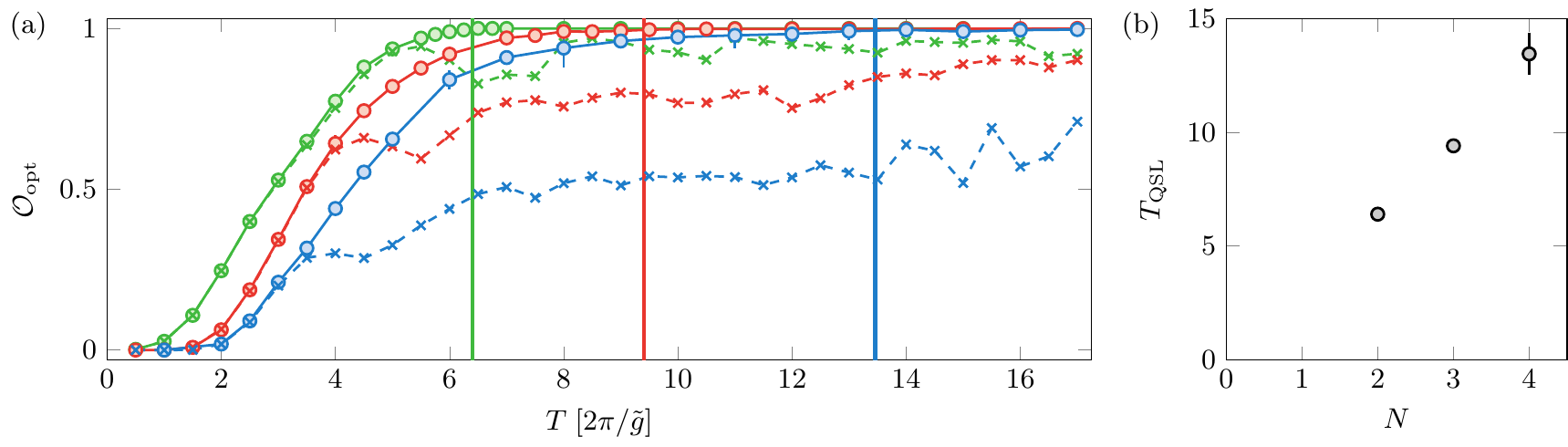}
\vspace{-6mm}
\caption{\label{Fig_Laughlin_optimized_ramp}
(a) Overlap between the Laughlin state and the state reached for an optimized ramp of duration $T$, either linear (crosses) or of arbitrary shape (dots), for atom numbers $N=2,3,4$ (green, red and blue, respectively). The vertical lines indicate the quantum speed limits $T_{\mathrm{QSL}}$. Error bars (most of them smaller than the point size) are calculated for the ramps of arbitrary shape, as the error from the extrapolation to an infinite number of time discretization steps. 
(b) Quantum speed limit $T_{\mathrm{QSL}}$ deduced from (a) as a function of the atom number $N$.
}
\end{figure*}
%%%%%%%%%%%%%%%%%%%%%%%%%%%%%%%%%%%%%%%%%%%%%

So far we considered linear detuning ramps of given speed $\dot\delta$, with initial and final detunings chosen far from resonance. In practice one may seek for detuning ramps of minimal duration leading to the Laughlin state with high fidelity. Minimizing the ramp duration could be crucial in experiments, where heating processes could prevent reaching the ground state for long ramps.  The issue of driving as fast as possible a quantum system into a given  state received a lot of attention in the recent years \cite{sklarz2002loading}, in particular in the context of quantum gate engineering. While the notion of quantum speed limit is well understood for time-independent Hamiltonians \cite{bhattacharyya1983quantum,margolus1998maximum,giovannetti2003quantum}, its extension to time-dependent systems was restricted to simple cases  \cite{anandan1990geometry,pfeifer1993fast,carlini2006time,rezakhani2009quantum,caneva2009optimal}. Optimal control of strongly-interacting systems brings up interesting open questions \cite{caneva2009optimal,doria2011optimal,rahmani2011optimal,caneva2011speeding,rahmani2013cooling,caneva2014complexity,van2015optimal}, such as the defect generation close to quantum phase transitions \cite{doria2011optimal,caneva2011speeding,van2015optimal}. 

Here we aim at maximizing the overlap $\mathcal{O}$ with the Laughlin state, obtained for a detuning ramp of fixed duration $T$, by adjusting the shape of the detuning $\delta(t)$. The Rabi coupling $\Omega$ is kept constant for simplicity. The optimization is performed by writing the detuning $\delta(t)$ as a function interpolating discrete values $ \{\delta_i(t_i)\}_{0\leq i\leq n}$, with $t_i$ uniformly spaced in the $[0,T]$ interval. The overlap is optimized over the $\delta_i$ values, using a stochastic variation algorithm  \cite{doria2011optimal}. The number  $n$ of discretization points is increased  up to $n=200$ steps, and the optimum overlap $\mathcal{O}_{\mathrm{opt}}$ is defined by extrapolating the calculated overlaps  to $n=\infty$ (see Appendix). Note that the optimized detuning ramps are highly irregular, similarly to the behavior of other physical systems \cite{rahmani2013cooling,van2015optimal,sorensen2016exploring} (see Appendix). The optimum overlap $\mathcal{O}_{\mathrm{opt}}$, calculated for atom numbers $N=2,3,4$ and ramp durations $0\leq T\leq 17\times2\pi/\tilde g$, increases monotonically with $T$  (see Fig.\,\ref{Fig_Laughlin_optimized_ramp}b).  We compare the performance of this optimized ramp with the overlap reached with a linear detuning ramp of same duration $T$. For a given duration $T$, we find the optimum start and end points of a linear ramp, maximizing the overlap value. As shown in Fig.\,\ref{Fig_Laughlin_optimized_ramp}a, using linear ramps results in much smaller overlap values, especially for long ramp durations and the largest atom number. 

In the case of detuning ramps of arbitrary shape, an optimum overlap consistent with 1 is reached for ramp durations $T$ larger than a threshold time $T_{\mathrm{QSL}}$ -- the quantum speed limit required to drive the system into the Laughlin state. We extract the duration  $T_{\mathrm{QSL}}$ by fitting the optimum overlap data with $\mathcal{O}_{\mathrm{opt}}>0.95$ using a piecewise linear function (see Appendix). As shown in Fig.\,\ref{Fig_Laughlin_optimized_ramp}b, the quantum speed limit $T_{\mathrm{QSL}}$ increases with the atom number. It would be interesting to calculate or measure experimentally the quantum speed limit for larger atom numbers, as it relates to the complex  many-body dynamics around a  topological critical point \cite{hamma2008adiabatic,caneva2011speeding}.

\section{Moore-Read state}

\begin{figure}
\includegraphics[width=\linewidth]{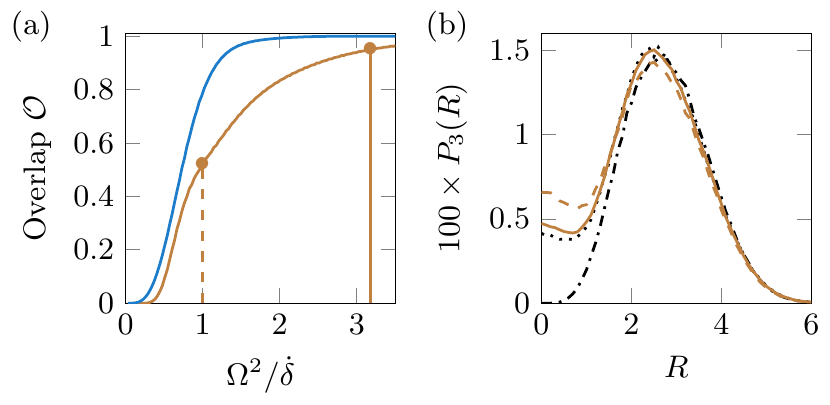}
\vspace{-6mm}
\caption{\label{Fig_MR}
(a) Overlap between the Moore-Read-like state $\ket{\psi_{\mathrm{MR}}^*}$ and the state reached for a detuning ramp of speed $\dot\delta$, for particle numbers $N=4,6$ (blue, brown lines, respectively), and a coupling $\Omega=\tilde g/(2\pi)$. The vertical lines indicate the two ramp speeds considered in (b).
(b) Density probability $P_3(R)$ for the 3-body hyper-radius $R$, calculated for the Moore-Read state $\ket{\psi_{\mathrm{MR}}}$, the actual ground state $\ket{\psi_{\mathrm{MR}}^*}$, and the states reached after detuning ramps of speed $\Omega^2/\dot\delta=1,3.2$ (dash-dotted and dotted black lines, dashed and solid brown lines, respectively), in the case $N=6$.
}
\end{figure}

We now consider the realization of the bosonic Moore-Read state, described by the wavefunction
\begin{equation}\label{eq_MR}
\psi_{\mathrm{MR}}=\prod_{1\leq i<j\leq N}(z_i-z_j)\,\mathrm{Pf}\left(\frac{1}{z_i-z_j}\right)_{1\leq i\neq j\leq N},
\end{equation}
where $\mathrm{Pf}$ denotes the Pfaffian of an anti-symmetric matrix \cite{moore1991nonabelions}. This state is
the analog for bosons of the Moore-Read state proposed to describe the FQHE at filling $5/2$ \cite{greiter1992paired}. It received a lot of attention due to the exotic nature of its elementary excitations, described as non-abelian anyons \cite{nayak2008non}. The ground state $\ket{\psi_{\mathrm{MR}}^*}$ of the LLL with contact interactions, of angular momentum $L=N(N-2)/2$ (for $N$ even), is expected to be close to the Moore-Read state $\ket{\psi_{\mathrm{MR}}}$ \cite{wilkin2000condensation}. %require ramp detuning, and showing high overlap values for ramp. %More physically than the many-body overlap, we investigate three-body correlations, in order to re
%Similarly to the Laughlin state, the Moore-Read state exhibits a flat density profile, reminiscent of its incompressibility in the thermodynamic limit. 
The Moore-Read state features a three-body anti-bunching, which can be revealed from the density distribution $P_3(R)$ of the 3-body hyperradius $R=\sqrt{|z_1-z_2|^2+|z_2-z_3|^2+|z_3-z_1|^2}$ ($P_3(R)$ being defined by analogy with the definition (\ref{eq_P2}) of $P_2(R)$). As shown in Fig.\,\ref{Fig_MR}b, the ground state $\ket{\psi_{\mathrm{MR}}^*}$ exhibits a significant three-body anti-bunching, yet with a non-zero value for $P_3(R=0)$.  Note that the exact realization of the Moore-Read state $\ket{\psi_{\mathrm{MR}}}$  would require implementing repulsive three-body interactions \cite{buchler2007three,roncaglia2010pfaffian}.

The Moore-Read-like state $\ket{\psi_{\mathrm{MR}}^*}$ can be realized in our scheme for an even particle number $N$, and a spin value $F=(N-2)/4$. We calculated numerically the state reached after a detuning ramp of finite speed, for particle numbers $N=4$ and $6$  (associated Hilbert spaces of dimension 20 and 2166, respectively). As shown in Fig.\,\ref{Fig_MR}a, the overlap between the prepared state and the actual LLL ground state is larger than 0.95 for ramp speeds $\dot\delta<0.3\,\Omega^2$ for $N=6$. In that regime, the distribution $P_3(R)$ of the 3-body hyperradius features a clear anti-bunching as $R\rightarrow0$ (see Fig.\,\ref{Fig_MR}b). These calculations show that FQHE-like states can be realized with atomic clusters using our method, provided the injection of angular momentum occurs on time scales comparable to the durations required for single-particle Landau-Zener adiabatic transitions. 

%%%%%%%%%%%%%%%%%%%%%%%%%%%%%%%%%%%%%%%%%%%%%
% Figure MR
%%%%%%%%%%%%%%%%%%%%%%%%%%%%%%%%%%%%%%%%%%%%%

%%%%%%%%%%%%%%%%%%%%%%%%%%%%%%%%%%%%%%%%%%%%%

\section{Trap ellipticity effects}

So far we restricted the discussion to the LLL $\ket{m;j,k=0}$, which is decoupled from $k\neq0$ states as long as the trapping potential is rotationally symmetric. We now discuss the impact of a trap anisotropy, corresponding to a harmonic confinement in the $x-y$ plane with trapping frequencies $\omega_x=\omega(1+\epsilon/2)$ along $x$ and  $\omega_y=\omega(1-\epsilon/2)$ along $y$. The anisotropy induces a coupling between states with different $k$ values, leading to a departure from the LLL. We consider here the effect of such a coupling on the generation of the Laughlin state $\ket{\psi_L}$. As the Hilbert space is significantly enlarged when considering the coupling to $k\neq 0$ states, we limit the analysis to $N=3$, for which the dimension of the Hilbert space to consider is 470 (it raises up to 15080 for $N=4$). 

 In the limit $\delta\rightarrow\infty$ and $\epsilon\rightarrow0$, we expect the ground state to be doubly degenerate, corresponding to two Laughlin states of opposite direction of rotation, constructed either within the states $\ket{m;j,k=0}$ (state $\ket{\psi_L}$) or $\ket{m;j=0,k}$ (state $\ket{\psi_L^*}$). A small anisotropy $\epsilon\neq 0$ induces a coupling between those states, leading to a strong departure of the actual ground state from the expected state $\ket{\psi_L}$. However, our scheme allows for an approximate realization of the Laughlin state $\ket{\psi_L}$, by keeping the detuning $\delta$ to a finite value. The Raman coupling then leads to a dressing of the single-particle quantum states, which breaks the symmetry between $\ket{m;j,k}$ and $\ket{m;k,j}$ states and favors the Laughlin state $\ket{\psi_L}$. In the regime of large detunings $\delta$, the effect of the Raman coupling can also be understood as an effective gauge field breaking time-reversal symmetry \cite{juzeliunas2005effective}.

\begin{figure}
\includegraphics[width=\linewidth]{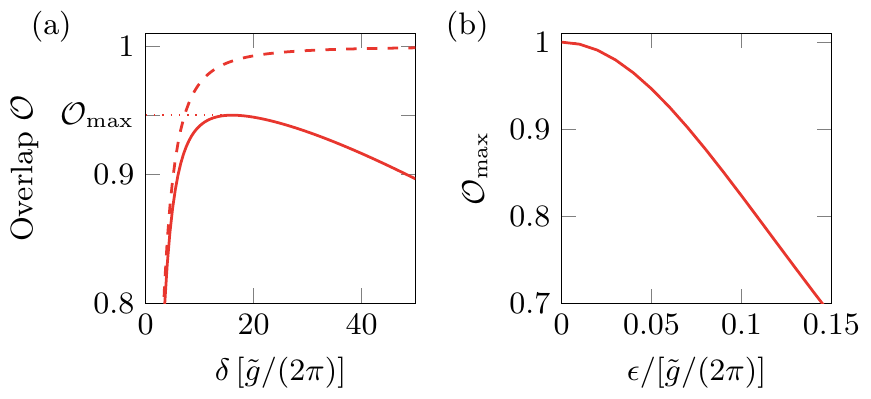}
\vspace{-6mm}
\caption{\label{Fig_Anisotropy}
(a) Overlap $\mathcal{O}$ between the Laughlin state $\ket{\psi_L}$ and the ground state of the full Hamiltonian for  $N=3$ (red line), a Raman coupling $\Omega=\tilde g/(2\pi)$, and  ellipticities $\epsilon=0$ (dashed line) and $\epsilon=0.05\,\tilde g/(2\pi)$ (solid line), as a function of the detuning $\delta$. A maximum value $\mathcal{O}_{\mathrm{max}}$ can be reached when varying the detuning $\delta$.
(b) Maximum overlap $\mathcal{O}_{\mathrm{max}}$ as a function of the ellipticity $\epsilon$.
}
\end{figure}

Taking into account all states $\ket{m;j,k}$, we calculated numerically the ground state for various values of $\epsilon$ and $\delta$, for the case $\Omega=\tilde g/(2\pi)$. For a perfectly isotropic trap ($\epsilon=0$), the overlap between the ground state and the Laughlin state $\ket{\psi_L}$ approaches 1 for large detunings $\delta\rightarrow\infty$ (see Fig.\,\ref{Fig_Anisotropy}a). For a non-zero trap ellipticity $\epsilon$,   balancing the residual population of $k>0$ states induced by the anisotropy and the $m_F=1/2$ states induced by the Raman coupling results in an optimal choice of $\delta$, leading to a maximal overlap $\mathcal{O}_{\mathrm{max}}$ (see Fig.\,\ref{Fig_Anisotropy}a). The maximal overlap  depends on the dimensionless parameter $\epsilon/\tilde g$  (see Fig.\,\ref{Fig_Anisotropy}b), and reaching $\mathcal{O}_{\mathrm{max}}>90\%$ requires anisotropies $\epsilon<0.07\,\tilde g/(2\pi)$. Using strong confinement along $z$ and/or Feshbach resonances,  interaction strengths  $\tilde g\sim 0.1$ to 1 can be obtained \cite{hadzibabic2006berezinskii,tung2010observation,hung2011observation,ha2013strongly}, leading to a constraint on the maximum allowed ellipticity in a small but achievable range $\epsilon\sim10^{-3}-10^{-2}$.

%%%%%%%%%%%%%%%%%%%%%%%%%%%%%%%%%%%%%%%%%%%%%
% Figure Anisotropy
%%%%%%%%%%%%%%%%%%%%%%%%%%%%%%%%%%%%%%%%%%%%%

%%%%%%%%%%%%%%%%%%%%%%%%%%%%%%%%%%%%%%%%%%%%%

\section{Conclusion}
We have presented a protocol for generating small clusters of atoms in FQHE states. It would be interesting to extend this work to fermionic  atoms, and to consider the effet of dipolar interactions, which should increase the stability of the Moore-Read state \cite{cooper2005vortex}, and lead to the formation of other exotic FQHE states \cite{cooper2005vortex,rezayi2005incompressible,cooper2007competing,seki2008incompressible,chung2008fermions}. While this proposal could be realized with most atomic species, Lanthanides such as Er or Dy would be most suited thanks to the large number of available spin levels, and the ability to apply Raman transitions with low residual heating due to spontaneous emission \cite{cui2013synthetic}. 

\section{Acknowledgements}
The authors would like to thank N. Regnault and J. Dalibard for fruitful discussions. This work is supported by the European Research Council
(Synergy grant UQUAM), the Idex PSL Research University (ANR-10-IDEX-0001-02), and R\'egion \^Ile de France (DIM NanoK, Atocirc project).

\section*{Appendix: Details on optimal control}
\begin{figure}
\includegraphics[width=\linewidth]{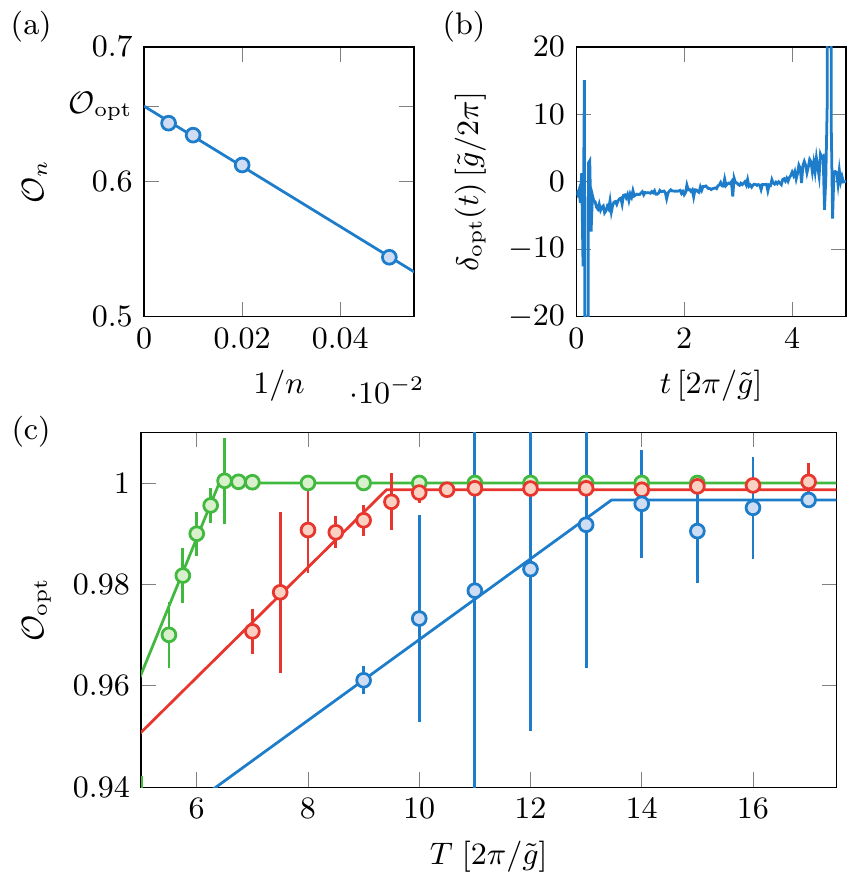}
\vspace{-6mm}
\caption{\label{Fig_Extrapolation}
(a) Optimum overlap $\mathcal{O}_n$ as a function of the inverse number of discretization steps $n^{-1}$, for $N=4$ and $T=5$. The optimum overlap $\mathcal{O}$ is obtained from the extrapolated value $\underset{n\rightarrow\infty}{\lim}\mathcal{O}_n$ , obtained from a linear fit of the data for $n\geq50$.
(b) Example of optimum detuning ramp $\delta_{\mathrm{opt}}(t)$ corresponding to $N=4$ and $T=5$, with $n=200$ discretization steps.
(c) Optimum overlaps $\mathcal{O}_{\mathrm{opt}}$ calculated for $N=2$, 3 and 4 (green, red and blue, respectively). The quantum speed limit times $T_{\mathrm{QSL}}$ are obtained from piecewise linear fits (solid lines).
}
\end{figure}

We provide additional details on the calculation of the optimum overlap with the Laughlin state that can be reached using a detuning ramp $\delta(t)$ of duration $T$. As discussed in the main text, the detuning ramp  is discretized in $n$ steps, and the optimization is performed on the discrete values $ \{\delta_i(t_i)\}_{0\leq i\leq n}$, with $t_i$ uniformly spaced in the $[0,T]$ interval. From the maximum overlaps $\mathcal{O}_n$ obtained using $n$ steps, we obtain the maximum overlap $\mathcal{O}_{\mathrm{opt}}$ as the value $\mathcal{O}_n$ extrapolated to $n=\infty$. An example of extrapolation is shown in Fig.\,\ref{Fig_Extrapolation}a.

We also show in Fig.\,\ref{Fig_Extrapolation}b an example of optimum detuning ramp $\delta_{\mathrm{opt}}(t)$, which reveals a typical irregular profile.

Finally, we discuss the fitting procedure to extract the quantum speed limit time $T_{\mathrm{QSL}}$ from the optimum overlaps $\mathcal{O}$. We use a piecewise linear function $\min[A,A+B(T-T_{\mathrm{QSL}})]$ to fit the data with $\mathcal{O}>0.95$, with $A$, $B$ and $T_{\mathrm{QSL}}$ as free parameters (see Fig.\,\ref{Fig_Extrapolation}c). 

%%%%%%%%%%%%%%%%%%%%%%%%%%%%%%%%%%%%%%%%%%%%%
% Figure Extrapolation
%%%%%%%%%%%%%%%%%%%%%%%%%%%%%%%%%%%%%%%%%%%%%

%%%%%%%%%%%%%%%%%%%%%%%%%%%%%%%%%%%%%%%%%%%%%

%\bibliographystyle{apsrev}
%\bibliography{references}

%\newpage
%
%\textbf{SUPPLEMENTARY INFORMATION}
%
%\section{Non-adiabatic Landau-Zener transition for higher-excited quantum states}
%In the main text we considered Landau-Zener transitions induced by a ramp of the detuning $\delta$ through 0. As the many-body ground state remains gapped for all values of $\delta$, 

\end{document}